\documentclass[12pt]{article}
\usepackage{epsfig}
\def\up#1{$^{#1}$}
\def\dn#1{$_{#1}$}
\newcommand{\A}{$A$}
\newcommand{\Z}{$Z$}
\newcommand{\etal}{{\it et al.}}
\newcommand{\hfthrds}{HF$\frac{2}{3}$S}
\newcommand{\bec}{\begin{center}}
\newcommand{\eec}{\end{center}}
\newcommand{\beq}{\begin{equation}}
\newcommand{\eeq}{\end{equation}}
\newcommand{\beqar}{\begin{eqnarray}}
\newcommand{\eeqar}{\end{eqnarray}}

\newcommand{\simlt}{\raisebox{-.6ex}{$\stackrel{\textstyle <}{\sim}$}}
\newcommand{\asi}{$\sim$}
\newcommand{\alf}{$\alpha$}
\title{\large \sf \begin{flushright} Report 
     NPI \v{R}e\v{z}--TH--03/2000
    \end{flushright}
\vspace{5mm}
 \bf On the reliability of the theoretical internal conversion 
     coefficients\thanks{Main part of this paper was presented at 
   the meeting of Decay Data Evaluation Project, Braunschweig, May 2000}}
\author{\normalsize  M.Ry\v sav\'y\thanks{e-mail: 
     rysavy@ujf.cas.cz}, O. Dragoun \\
\small \it Nuclear Physics Institute, Acad. Sci. of Czech Republic,
    \vspace{-0.7ex}\\
\small \it CZ-250 68 \v Re\v z near Prague, Czech Republic}
\date{ }
\begin{document}

\maketitle
\title{}
%%%%%%%%%%%%%%%%%%%%%%%%%%%%%%%%%%%%%%%%%%%%%%%%%%%%%%%%
\begin{abstract}
Possible sources of uncertainties in the calculations of the internal
 conversion coefficients
 are studied. The uncertainties induced by them are estimated.
\end{abstract}
%%%%%%%%%%%%%%%%%%%%%%%%%%%%%%%%%%%%%%%%%%%%%%
%%%%%%%%%%%%%%%%%%%%%%%%%%%%%%%%%%%%%%%%%%%%%%%%%%%%%%%%
\section{Introduction}
The internal conversion coefficients (ICC) serve many years as an
indispensable tool 
to assign spin--parity quantum numbers of excited nuclear levels.
There exist widely distributed
 tables \cite{Roe78,Ban78,Hag68} of ICC covering
broad region of transition energies,  atomic numbers 
from \Z=30 \cite{Roe78,Hag68} or 10 \cite{Ban78}
up to \Z=103 \cite{Hag68} or \Z=104 \cite{Roe78,Ban78} and
transition multipolarities M1 -- M4, E1 -- E4.
Very recently, tables of ICC for the superheavy elements 
(104$\le$Z$\le$126) appeared \cite{Rys00}.

When computing ICC, however, various approximations are applied.
Some of them are of purely theoretical origin, e.g. the calculations
are done only for the first non-vanishing order of the perturbation
theory. Other ones are connected with improper description of
physical reality, e.g. the ICC is calculated for a free atom whereas
the real source atom is usually chemically bound in some compound
or embedded into a solid.
In this work we systematize these particular sources
of uncertainties and estimate their effect on the accuracy
of the calculated ICC.
\section{Basic formulae}
The basic formulae for the ICC together with its derivation
may be found in many works. We present here those from the monography
\cite{Anom}.
The ICC for the transition of the multipolarity $\tau L$ 
and the atomic shell $i$ is given by
\beq
\label{e:ICC}
\alpha_i(\tau{L}) = \alpha \pi \omega \frac{2j_i + 1}{L(L + 1)}
 \sum_{\kappa_f}{| C^{j_f,-1/2}_{j_i,-1/2,L,0} ({\cal R}^{(\tau L)}_{\kappa_f}
 + {\cal T}^{(\tau L)}_{\kappa_f}) |^2}.
\eeq
Here
\beqar
\label{e:rML}
 {\cal R}^{(ML)}_\kappa = \int_0^\infty{{\cal F}^{(0)}_\kappa(r) 
   h_L(\omega r) dr}, \\
\label{e:rEL}
{\cal R}^{(EL)}_\kappa = \int_0^\infty{[ L {\cal F}_\kappa(r) 
   h_L(\omega r) + {\cal F}^{(-1)}_\kappa(r) h_{L-1}(\omega r) ] dr}.
\eeqar
and
\beqar
\label{e:f0}
{\cal F}^{(0)}_{\kappa_f}(r) = (\kappa_i + \kappa_f) (u_{\kappa_i}
   v_{\kappa_f} + u_{\kappa_f} v_{\kappa_i}),\\
\label{e:f}
{\cal F}_{\kappa_f}(r) =  (u_{\kappa_i} u_{\kappa_f} +
    v_{\kappa_i} v_{\kappa_f}),\\
\label{e:fminus1}
{\cal F}^{(-1)}_{\kappa_f}(r) = (\kappa_i - \kappa_f - L) v_{\kappa_i}
   u_{\kappa_f} + (\kappa_i - \kappa_f + L) u_{\kappa_i} v_{\kappa_f}.
\eeqar
%%%%%%%%%%%%%%%%%%%%%%%%%%
$C^{j,m}_{j_1,m_1,j_2,m_2}$ is Clebsch-Gordan coefficient, \alf\ is 
the fine structure constant, $h_l(x)$ is the spherical Hankel function, 
$\omega$ is transition energy, $u_\kappa$ and $v_\kappa$ are small
and great, respectively, component of the solution of Dirac equation
for electron.

For the transition of magnetic multipolarity, the quantity
\beq
{\cal T}^{(ML)}_{\kappa_f} = \frac{\int{d^3\!R \ \vec{J}_{nucl}(\vec{R})\ 
  \hat{\vec{L}}\  g^{(3)}_{\kappa_f}(R)\  Y^*_{LM}}}
   {\int{d^3\!R \ \vec{J}_{nucl}(\vec{R})\  \hat{\vec{L}}\  j_L(\omega R)\ 
    Y^*_{LM}}}
\label{e:Tig}
\eeq
where
\beq
g^{(3)}_\kappa(R) = h_L(\omega R) \int_0^R{dr \ j_L(\omega r) \ 
   {\cal F}_\kappa(r)} - j_L(\omega R) \int_0^R{dr \ h_L(\omega r)
   {\cal F}_\kappa(r)}
\eeq   
is the only one which containes nuclear variables. It describes
``intranuclear conversion'' (see  Sect. \ref{s:intra}). For
the electric transitions, the analogical quantity 
${\cal T}^{(EL)}_{\kappa_f}$ is given by a more complicated 
formula but it has similar structure.

From the formulae above we see that the ICC depends on four `parameters':
transition energy $\omega$, transition multipolarity $\tau L$, 
initial electron state (i.e. atomic shell), and atomic number $Z$.
For the atomic number, the dependence is purely implicit -- via
atomic potential which then forms the electron wave functions 
$u_\kappa$ and $v_\kappa$. The other quantities enter into the formulae
explicitly but, moreover, they also affect the result indirectly. E.g.
the transition multipolarity determines the allowed final electron
states due to the conservation of angular momentum. It is obvious that
to draw some  information on the shape of the 
dependencies directly from the above formulae is almost impossible.
The only way is to evaluate ICC for various parameters and find out
the dependencies from the numerical results. 
Analogically, the effect of various approximations and/or model
imprefections may also be studied via numerical calculations.
\section{Sources of the ICC uncertainties}
When we put a question about the precision of the theoretical ICC
we must first specify the possible sources of their uncertainties.
 These may be
very roughly divided into two classes --- those stemming from theory and
those given by our insufficient knowledge of some quantities entering the 
calculations. Into the first category there can be included {\em atomic
model}, {\em nuclear deformation and nonsphericity in general}, and
{\em higher order effects}. The second one consists of {\em atomic
binding energies}, {\em isotopic effect}, and {\em chemical effects}.
Somewhere in-between is the question of correct accounting of
{\em nuclear structure}, i.e. the {\em intranuclear conversion}.
And, finally, a serious error may be introduced by incorrect
interpolation in existing tables, especially in case of total
ICC.

In the following, we address ourselves to the particular items 
listed above. We summarize all the known facts and try to give
a realistic estimate of the  uncertainties induced.
\subsection{Atomic model}
\label{s:atomic}
As seen from Eqs. (\ref{e:f0} - \ref{e:fminus1}), the electron
wave functions ($u_\kappa$ and $v_\kappa$) are needed to calculate
an ICC. The evaluation of these wave functions is based on some
suitable model of atom.

When we skip the `ancient' calculations with the model of point nucleus
without screening, the first more realistic ICC values for the K and L shells
were obtained
\cite{Ros58}-\cite{Sli58} using the Thomas-Fermi-Dirac atomic model.
Later on this model was abandoned in favour to physically more founded
model of Hartree and Fock with the Slater exchange
term (HFS). This atomic model was employed to provide the first 
tables of the N-shell ICC \cite{Dra69a}.
A little sooner in Ref. \cite{Hag68}, a modified HFS method where the
Slater exchange term is weighted by a factor of 2/3 (so called
\hfthrds\ method) was utilized to get the ICC for the shells
K, L, and M. 

In the late 70', the most extensive tables \cite{Roe78,Ban78}
appeared where the pure HFS model was applied.
Approximately at the same time the first ICC based on the true
 Hartree--Fock (HF) model 
(without the Slater exchange term) were calculated \cite{Dra77,Dra81}.
To complicate the things yet more, it is not clear whether the `hole'
in the atomic shell remaining after the conversion electron should be 
taken into account or not.

Evidently one should hesitate which ICC are the ``real'' ones and 
what error can be made by using other ones.
Band et al. \cite{Ban81} performed extensive calculations of ICC
in both models HFS and \hfthrds\ and for both variants `hole' and
`no hole'. They evaluated large sets of ICC for \Z=30, 60, 90,
 conversion electron 
energy between 10 and 400 keV, multipolarities M1 -- M4 and E1 -- E4,
and K, L\dn{1}, L\dn{2} and L\dn{3} subshells.
Then they constructed the quantities which may characterize 
the uncertainty caused by the atomic model,
\beq
\Delta_1 = [\frac{ICC(hole)}{ICC(no\ hole)} -1] \times 100,
\eeq
\beq
\Delta_2 = [\frac{ICC(HF\frac{2}{3}S)}{ICC(HFS)} -1] \times 100,
\eeq
where the symbol in parentheses indicates the model used. 
These quantities vary with multipolarity -- in general, their
absolute values increase  with $L$. They increase with 
{\em decreasing} energy.
For some combinations of subshell and multipolarity they do not
exceed 0.5\%, for the remaining ones there are graphs in \cite{Ban81}.

In {\cite{Dra81}, an extensive comparison of HF 
model\footnote{In fact, only the bound electron wavefunctions are
calculated in HF model there. The wavefunctions of the emitted
electrons are generated as previously is HFS model. It was 
proved that the free electron wavefunctions are much less
sensitive to the model details (e.g. \cite{Zil81}).} 
ICC with
experimental ones was performed. Altogether 57 experimental data items for
8  transitions in six different isotopes were examined.
A sensitive statistical test revealed better agreement of the theory with
experiment for the HF ICC than for the HFS ones.
The differences between the two sets of theoretical ICC varied from
$\simlt$1~\% on the K-shell, 2--3~\% on the L-subshells
up to $\sim$30~\% on the outermost shell.

More recently, there appeared \cite{Ban90} the `true' HF model ICC, i.e.
those where even the free-electron wave functions were evaluated without
the approximation of the exchange term. 15 transitions in various
isotopes were studied and better agreement of the theory and
experiment was confirmed.

Tables of ICC in the HF model are not available at present. However,
as stated above, programs exist and it is possible to calculate
the HF (i.e. physically best) ICC for particular cases.
As for their uncertainty, only that corresponding to numerical
accuracy (mostly $<$0.5~\%) may be expected. For the other
atomic models, it is  difficult to give some general error 
estimate. It might be the difference from the HF ICC. Anyway
these quantities are known for several particular transitions
only (e.g. \cite{Dra81,Ban90}).
 And in any case, such an `error' would be a systematic
rather than statistical one.

{\it \underline{Recommendation}}: For  analysis of
precise experimental data, use the HF model based ICC.
%%%%%%%%%%%%%%%%%%%%%%%%%%%%%%%%%%%%%%%%%%%%%%%%%%%%%%%%%%%%%%%
\subsection{Nuclear deformation and nonsphericity}
The formulae for conversion coefficients are based on the assumption
that the atomic field is spherically symmetrical. 

It has been proved long time ago \cite{Cou66}
that the closed shells form a symetrical atomic field.
With several exceptions (\cite{Cas98}, p. 74), there is only one unclosed
 shell in every free atom. 
This is  the valence shell which, in
most cases, is the outermost one. (Even if there are two unclosed
shells, they always belong to the outer ones.) The calculated
ICC is in fact the mean over all projections of the angular momentum.
This evidently brings an error in the case of unclosed shells.
If the actual occupation number for the particular shell, $i$,
 is $n_{act}$
and the maximum possible one is $n_0$, the conversion coefficient 
should be rescaled as
\beq
\label{e:scale}
      \alpha_i(act) = \frac{n_{act}}{n_0} \times \alpha_i(calc),
\eeq
where $\alpha_i(calc)$ is the calculated ICC.
It greatly reduces the error caused by the absence of one 
or several electrons.
The error due to the asymmetry remains but it can be negtected.
The asymmetry is in fact a slight perturbation which affects 
almost exclusively only
 the particular unclosed shell. As stated above, it concerns the
outer shells and there is a greater uncertainty due to the 
chemical environment (Sect. \ref{s:envir}).

Another possible source of asymmetry can be a deformed atomic
nucleus. The effect of nuclear deformation on the ICC was
studied in \cite{Mat68,Bor74,Chu60}.
The authors \cite{Mat68,Bor74} assumed a nucleus with high 
quadrupole deformation and used
perturbation series to compute contributions to ICC. Such
calculations were done  for $^{160}_{\ 66}$Dy, 
$^{166}_{\ 68}$Er, $^{170}_{\ 70}$Yb, and $^{186}_{\ 76}$Os
nuclei exhibiting high quadrupole moments. The comparison of
the results with the ICC evaluated in a usual way (i.e. for spherical
nucleus) revealed a negligible change of $\sim$10\up{-2}~\%.
 In \cite{Chu60}, also the effect of magnetic dipole
moment was investigated and  found to be negligible.

It should be kept in mind that the transitions in
deformed nuclei  are often hindered. In such  cases
it may be necessary to account for the intranuclear conversion
(Sect. \ref{s:intra}).

{\it \underline{Recommendation}}: If using the conversion coefficient
for a subshell which is not completely filled do not forget
to rescale it following Eq.(\ref{e:scale}). If studying a
transition in a deformed nucleus, think on possibility of
intranuclear conversion.
%%%%%%%%%%%%%%%%%%%%%%%%%%%%%%%%%%%%%%%%%%%%%%%%%%%%%%%%%%%%%%%%
\subsection{Higher order effects}
\label{s:higher}
At present, all ICC are evaluated up to the first non-vanishing order
of perturbation theory, i.e. the first order for the gamma 
emission and second one for the emision of conversion electron. The 
Feynman diagrams corresponding to the next nearest orders are
presented in Fig.\ref{f:Fey}. The contributions of individual
graphs to ICC were estimated by several authors, e.g.
\cite{Kru58}--\cite{Hin81},\cite{Ban90}.

\hspace*{-\parindent}
\parbox{11cm}{\epsfxsize=9cm \epsfysize=7cm \epsfbox[20 500 330 810]
{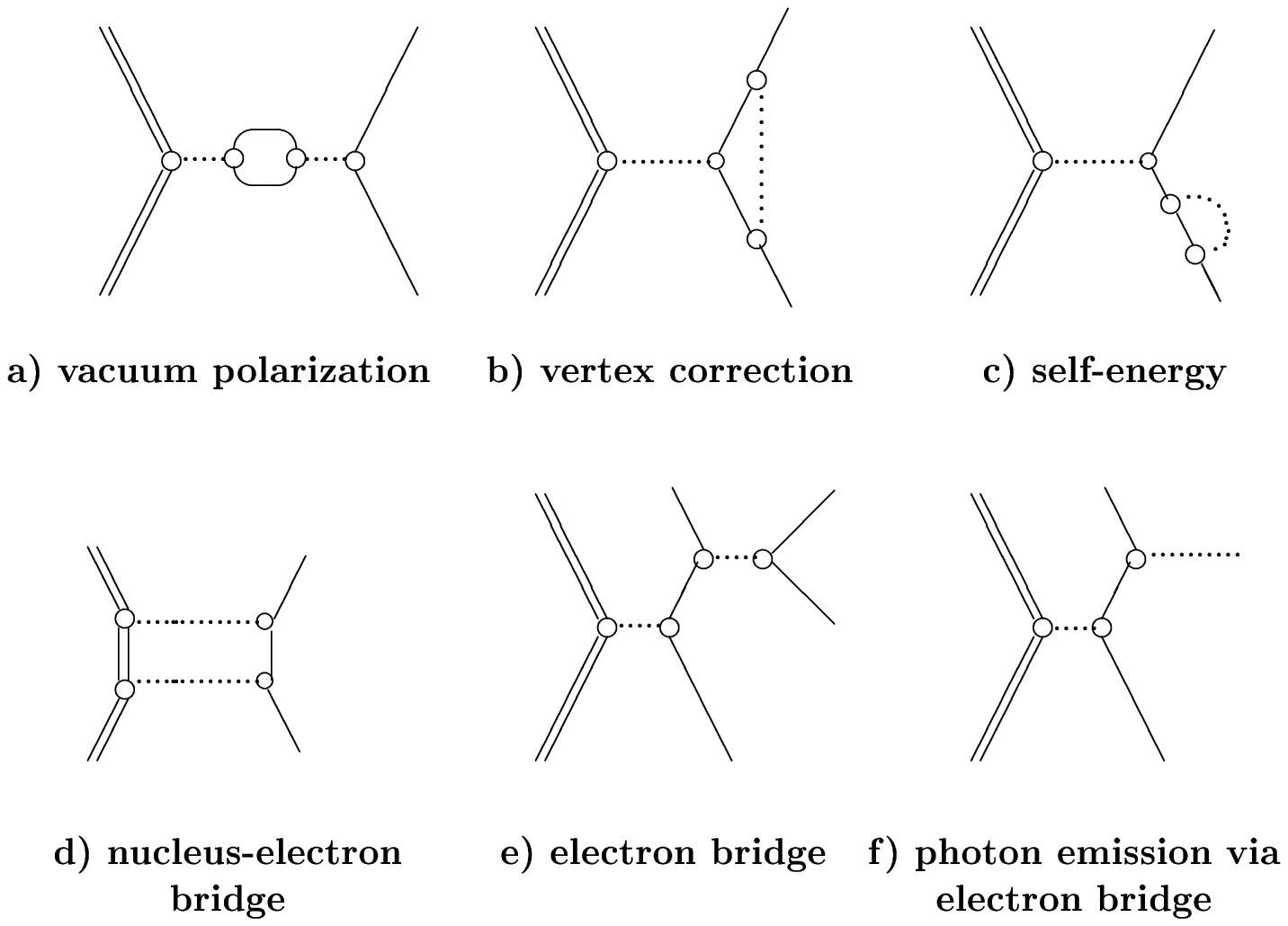}} \vspace{4ex}
\begin{figure}[h]
\protect\caption[Fig1]{Feynman diagrams for internal conversion up to the
  fourth order and for $\gamma$--ray emision up to the third order of 
perturbation theory.}
\label{f:Fey}
\end{figure}

\underline{Vacuum polarization.} It is the only higher order effect
which can be taken into acount precisely by a change of atomic potential.
It was studied in \cite{Raf72}. The authors found small
influence on ICC, namely a change of $<$0.1~\% for \Z=26 and
$<$1~\% for \Z=80.

\underline{Vertex correction.} Contribution of vertex correction
to ICC was never investigated. Pauli \etal\ \cite{Pau75} 
stated that Hager and Seltzer \cite{Hag70} studied
it.
However, the graphs presented in \cite{Hag70} correspond to `electron
bridge' -- graphs (e) and (f) in Fig.\ref{f:Fey}.

\underline{Self--energy.} Up to our knowledge the possible
contribution of this term to the ICC was never examined.

\underline{Bridges.} These corrections were extensively investigated
by Krutov \etal\ \cite{Kru58}--\cite{Kru90}. Unfortunately, there are
almost none numerical estimates. The up to 20~\% effect \cite{Kru70}
of the graph (d) in one specified case seems unrealistic 
(during the calculations, a lot
of terms which can contribute were neglected). Hager and Seltzer
\cite{Hag70} stated about 5~\% efect of (e) and (f) on the
L-subshell ICC in $^{182}_{\ 74}$W,
$^{169}_{\ 69}$Tm, and $^{160}_{\ 66}$Dy. Also here, several important
terms were neglected and, moreover, the Pauli principle
was not fulfilled in their calculations. 
On the other hand, Mayol \etal\ \cite{May84} found the electronic bridge
effect of 0.0047~\% in \dn{77}Ir atom which is stated to be the maximum 
one from 11 isotopes studied.
Hinneburg \etal\ \cite{Hin77} reported extremely
small contribution of 10\up{-9}.
In the later works \cite{Hin79,Hin81} they concentrate to the
transition with the energy of 77~eV in $^{235m}_{\ \ \ 92}$U where
the graph (f) could raise the photon decay constant by a factor of
$\sim$10\up{5} and thus suppress the ICC. Their calculations,
however, cannot be verified since the ICC of such a low
energy transition is unmeasurable at present. Band \etal\ \cite{Ban90} 
admit
the effect of `electron bridge' for very low transition energies but
do not present any concrete estimates.

Anyway, the above approach  -- the evaluation of contributions
of individual graphs -- is not correct. The right method is
first to derive the {\em matrix elements} corresponding to 
particular graphs and add them together. Only then the modified
integrals (\ref{e:rML},\ref{e:rEL},\ref{e:Tig}) can be applied in 
the final formula (\ref{e:ICC}).
This is such a difficult task that 
nobody even dared to perform it up to now. 

{\it \underline{Recommendation}}: For the transition energies not too
close to conversion threshold, assume the uncertainty caused by
neglection of higher orders to be about 1~\%. This estimate follows
 from analogy with electron Coulomb scattering using the
formulae in \cite{Gas66}. As for the very low transition energies,
we can say nothing at present.
%%%%%%%%%%%%%%%%%%%%%%%%%%%%%%%%%%%%%%%%%%%%%%%%%%%%%%%%%%%%
\subsection{Atomic binding energies}
\label{s:bind}
The binding energies, $E_{b,i}$, of the atomic electrons enter into 
the ICC via the relation
\beq
\label{e:bineng}
   E_{kin,i} = E_{tr} - E_{b,i},
\eeq
where $E_{kin,i}$ and $E_{tr}$ are kinetic energy of conversion electron
and transition energy, respectively. In Eq.(\ref{e:bineng}), the 
recoil energy as well as work function of the source material are
neglected. The $E_{kin,i}$ specifies the conversion electron wave 
functions appearing in formulae 
(\ref{e:f0}--\ref{e:fminus1}).

The binding energies are presented in \cite{Sev74} with the 
typical error of $\pm$1~eV, sometimes 2 or 3~eV. (An 
exception are several transuranic elements where the stated
error amounts in some cases up to $\pm$18 eV.)
To estimate the uncertainty in ICC caused by those uncertainties
in $E_{b,i}$ we
performed calculations with various values of binding energies.
In particular, we evaluated ICC for \dn{26}Fe, \dn{55}Cs and
\dn{87}Fr and the subshells (K,M\dn{3}), (K,N\dn{5}) and (K,O\dn{5}),
respectively. This setting was chosen to cover low, medium and
high Z and always one inner and one outer shell. The transition
energies of 30, 50, 100, 200, 300 and 500~keV were used where applicable.
Finally, the ICC were evaluated always for three binding energies,
$E_0$-10~eV, $E_0$, and $E_0$+10~eV where $E_0$ is the
binding energy \cite{Sev74}. (Note that the binding energies --
including those for the inner shells --  may
also be influenced by the chemical state of the converting atom.
These so called {\em chemical shifts}, however, are rather small
and our interval of 20 eV covers them all right.)
\begin{table}[h]
\caption{\protect  %\small
Internal conversion coefficients for \dn{55}Cs, shell K,
transition energy of 50~keV  and three binding energies.
}
\label{t:bind}  \vspace{2ex}
%\begin{center}
\begin{tabular}{clllr}
\hline
mult. &  $E_0$-10~eV  & $E_0$\up{a} & $E_0$+10~eV &   $\Delta$\up{b} \\
\hline
  E1 &   1.2320(0)\up{c} &  1.2318(0) & 1.2317(0)  &  .02\\
  E2 &   7.0612(0) &  7.0563(0) & 7.0513(0)  &  .07\\
  E3 &   2.6671(1) &  2.6632(1) & 2.6593(1)  &  .29\\
  E4 &   9.2835(1) &  9.2622(1) & 9.2408(1)  &  .46\\
  M1 &   5.8912(0) &  5.8912(0) & 5.8911(0)  &  $<$.01\\
  M2 &   9.4640(1) &  9.4617(1) & 9.4594(1)  &   .05\\
  M3 &   6.1914(2) &  6.1863(2) & 6.1811(2)  &   .17\\
  M4 &   3.1418(3) &  3.1369(3) & 3.1320(3)  &   .31\\
\hline
\end{tabular}\\
\up{a} $E_0$=36189.9 eV\\
\up{b} Maximum change in percent, $\Delta =
        (\frac{max}{min} - 1)\times 100$\\
\up{c} x(y) means x$\times$10\up{y}
%\end{center}
\end{table}

As a measure of the uncertainty we take the quantity $\Delta =
(\frac{max}{min} - 1)\times 100$ where $max$ and $min$ are
the maximum and minimum, respectively, ICC from the three ones 
corresponding to different binding energies. 
For the innermost (K) shell, the results are following.
In almost all cases, $\Delta$ is of order of 10\up{-2} of percent,
up to 0.1~\%. The exception is the 50~keV transition
in \dn{55}Cs (see Tab.\ref{t:bind}). For  high multipolarities,
the changes there reach almost 0.5~\%. The reason is simple -- we
are closer to threshold than in the other cases. (When approaching
the threshold, the kinetic energy decreases and its change plays
more important role.) For the outer shells, the changes are even
less.

The change of binding energies used here (20~eV) is relatively
large. If the experimental error of binding energy is less
(which is usually the case), the corresponding uncertainty in
ICC is also less.

{\it \underline{Recommendation}}: If the binding energy is known
with a precision better than say $\pm$5~eV and the transition energy
is high enough (20~keV or more above the threshold), the 
uncertainties caused by it in ICC are negligible. This is true
even when we keep in mind the possible chemical shifts
of binding energies.
For the transition energy very close to threshold, no general 
rule can be given and the problem must be solved by
calculating ICC for different binding energies.
%%%%%%%%%%%%%%%%%%%%%%%%%%%%%%%%%%%%%%%%%%%%%%%%%%%%%%%%%%
\subsection{Isotopic effect}
\label{s:isot}
The mass number, \A, does not enter  the formulae for ICC
explicitly. Nevertheless it determines the nuclear radius
\beq
\label{e:rnucl}
      r_{nucl} = r_0 \times A^{1/3}.
\eeq
This implies that \A\ affects the atomic potential which determines
the electron wave functions. 
We describe the atomic nucleus by the Fermi distribution 
of charge
\beq
     \rho(r) = \frac{\rho_0}{1 + \exp{(\frac{c - r}{t})}},
\eeq
where $c$ and $r_{nucl}$ are tabulated in \cite{Lu71} and $t$ is
found from the condition $\rho(r_{nucl}) = 0.05\rho_0$.

We assume that $r_{nucl}$ in \cite{Lu71} corresponds to
 \A\ of the most abundant isotope for every element. 
To obtain ICC accordant with 
another mass number, $A'$, we rescaled the nuclear radius by
the ratio
$(A'/A)^{1/3}$.
The calculations were performed for the same set of elements,
transition energies, shells and multipolarities as that
in Sect.(\ref{s:bind}). ICC were evaluated for
three different mass numbers, \A\ and $A\pm 10$, in $_{55}$Cs
(\A=133) and $_{87}$Fr (\A=223) and for two mass numbers (57 and 67)
in the case of $_{26}$Fe.

The results are following: In iron and cesium, only very small
changes (up to $\sim$0.2~\%) were observed for the K-shell ICC.
For the outer subshells considered, the changes were significantly lower.
In the francium, the changes turned out to be higher 
(see Tab.\ref{t:isot}).
\begin{table}[h]
\caption{\protect  %\small
Internal conversion coefficients for \dn{87}Fr, shell K,
transition energy of 200~keV  and three mass numbers.
}
\label{t:isot}  \vspace{2ex}
%\begin{center}
\begin{tabular}{clllr}
\hline
mult. &  \A=213  & \A=223 & \A=233 & $\Delta$\up{a} \\
\hline
  E1  &  7.0255(-2)\up{a} &  7.0251(-2) &  7.0250(-2) &    0.01\\
  E2  &  1.6199(-1) &  1.6197(-1) &   1.6196(-1) &    0.02\\
  E3  &  3.8503(-1) &  3.8490(-1) &   3.8473(-1) &    0.08\\
  E4  &  9.6301(-1) &  9.6236(-1) &  9.6176(-1) &    0.13\\
  M1  &  1.7636(0) &   1.7586(0)  &   1.7538(0) &   0.56\\
  M2  &  6.6561(0) &   6.6465(0)  &  6.6379(0)  &    0.27\\
  M3  &  1.7317(1) &  1.7295(1)  &   1.7277(1) &   0.23\\
  M4  &  4.2730(1) &   4.2689(1) &   4.2642(1) &    0.21\\
\hline
\end{tabular}\\
\up{a} See comments on Tab.\ref{t:bind}
%\end{center}
\end{table}
Generally, the changes do not depend to transition energy. This
indicates that they are caused mostly by slight alterations of
the bound--electron wave functions. As could be expected,
the K-shell is affected most of all. Moreover, the magnetic
transitions are more sensitive than the electric ones.

{\it \underline{Recommendation}}: In most cases, the uncertainties 
caused by isotopic effect are negligible. In case that data on
some extremely heavy or extremely light isotop of a given
element are studied, allow for uncertainty of up to 0.5~\%
to be on the safe side. A better alternative is to evaluate
ICC directly for the given \A\ which eliminates this kind
of uncertainties completely.
%%%%%%%%%%%%%%%%%%%%%%%%%%%%%%%%%%%%%%%%%%%%%%%%%%%%%%%%%%%%%%%
\subsection{Chemical effects}
\label{s:envir}
The conversion coefficients are always evaluated under the
assumption that the source atom is an isolated free neutral atom.
In practice, the experimental ICC are obtained from a 
radioactive source where the radiating atoms are bound in some
structure -- molecule or crystal. 

There were attempts to account for the source structure.
The Wigner--Seitz boundary condition introducing solid state
effects  was applied in calculations \cite{May84}. The ICC were
evaluated for 11 transitions in various isotopes. It turned out
that the K-shell and total ICC were almost the same as those evaluated
in the free atom approximation. (As for the other shells, no information
is given in \cite{May84}.) In \cite{Ban84}, the ICC for the free atom
were compared with those calculated using the X$\alpha$-SW
description of a Tc compound. The difference between the two total ICC
was of order of 10\up{-1}~\%.
In \cite{Har83}, the X$\alpha$-SW calculations for some tellurium compounds 
gave electron densities near the Te nucleus. The free-atom ICC
were then scaled as
ICC($real$) = $|\psi^{X\alpha-MS}(r_{nucl})|^2/|\psi^{free}(r_{nucl})|^2
\times$ ICC($free$), where $|\psi|^2$'s are the bound electron
 densities at
the nuclear edge. Unfortunately this attepmt failed to bring
the theoretical ICC to an agreement with experimental ones
for several Te compounds.

Another possibility is to approximate the `real' ICC by their
values evaluated for free ions. The atom bound e.g. in a molecule
certainly is {\em not} a free ion. In spite of this the  ICC
for free ions were calculated \cite{Bra82}-\cite{Vse90} (for 
$^{125}_{\ 52}$Te, $^{83}_{36}$Kr, $^{206}_{\ 83}$Bi, and
$^{142}_{\ 59}$Pr, respectively) and it turned out that 
the experimental outer--shell ICC ratios lay within the range
of theoretical values for reasonably ionized atoms.
This gives us an argument to use the dispersion of ICC calculated
for several ionic states as a measure of uncertainty caused by
the chemical bond.

First, we performed extensive calculation of ICC in 
$^{133}_{\ 55}$Cs over a broad region of ionizations (from -1
up to +9). The multipolarities M1 -- M4 and E1 -- E4, all atomic shells,
and the energies of 30, 100, 200, and 500 keV were treated.
The method of calculation is described in \cite{Rys92}.

%
%\hspace*{-\parindent}
\parbox{11cm}{\epsfxsize=9cm \epsfysize=7cm \epsfbox[20 90 330 410]
{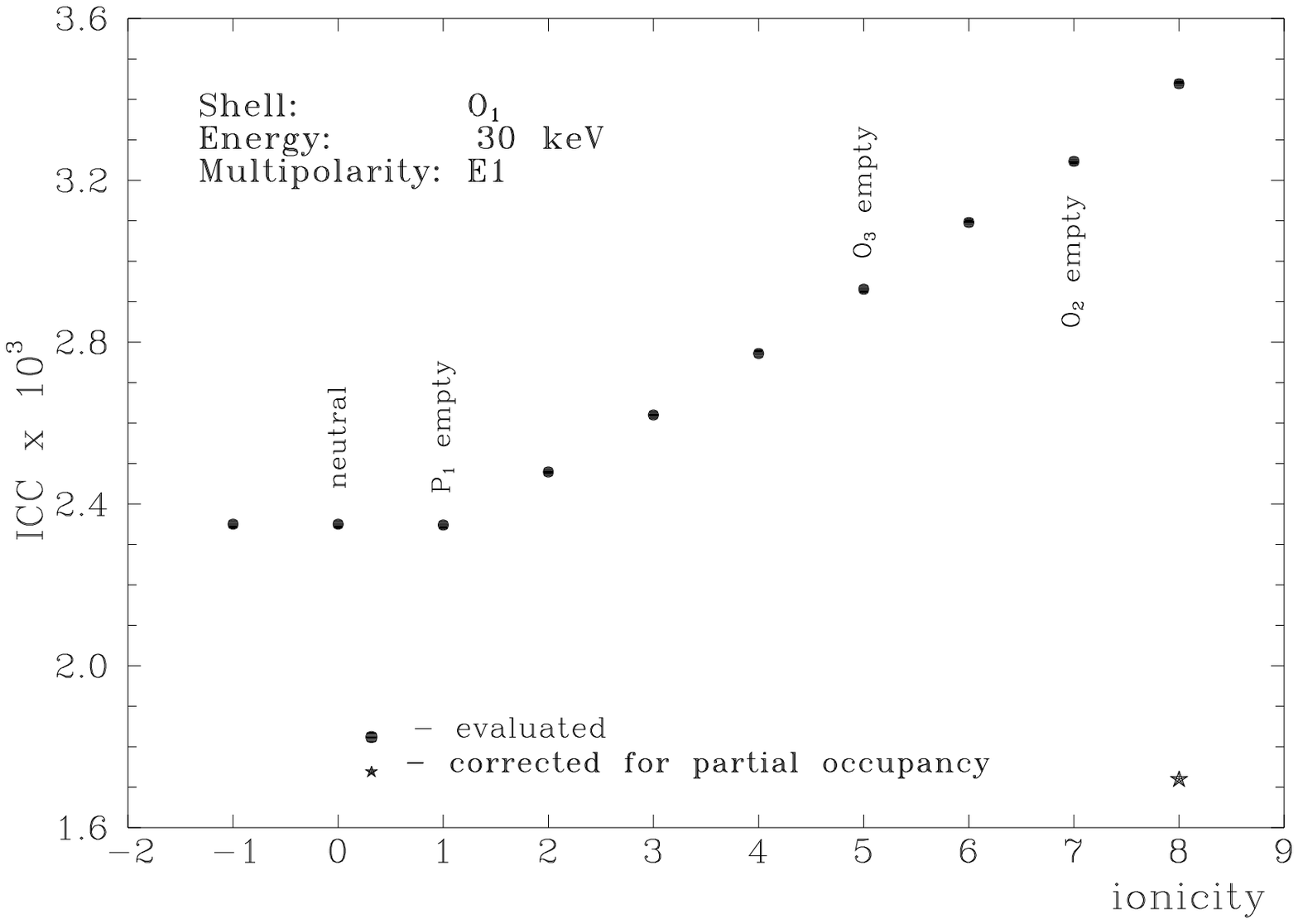}} \vspace{4ex}
\begin{figure}[h]
\protect\caption[Fig2]{ICC on the O\dn{1}-subshell of \dn{55}Cs
in dependence of ionization of the atom. With the exception 
of the ionization +8, the O\dn{1}-subshell is fully occupied.}
\label{f:chem}
\end{figure}

As a measure of uncertainty we used again the quantity $\Delta$ 
introduced in Sect.\ref{s:bind} (Tab.\ref{t:bind}) but now the
maximum and minimum is always taken over all {\em ionizations}.
The magnitudes of $\Delta$ were as follows: $\simlt$0.2~\% (for
the shells K, L\dn{1}--L\dn{3}, M\dn{1}--M\dn{3}), $\sim$0.3~\%
(M\dn{4}, M\dn{5}), $\sim$0.6~\% (N\dn{1}), increasing up to 10~\%
(for N\dn{5}), and several tens of percent (O-subshells).
For the total ICC, $\Delta$ was mostly $\simlt$0.3~\%, exceptionally
$\sim$0.5~\% (for M1 transition, energy 30 keV).
The results are not too dependent on both energy and multipolarity.
For illustration there are results for the subshell M\dn{2} in
Tab.\ref{t:chem}. The changes of ICC on the O\dn{1}
shell shows Fig.\ref{f:chem}.

\begin{table}[h]
\caption{\protect  %\small
Uncertainties (in percent) of ICC on the M\dn{2}-subshell
of $^{133}_{\ 55}$Cs caused by chemical binding.}
\label{t:chem}  \vspace{2ex}
%\begin{center}
\begin{tabular}{cllll}
\hline
 \  & \multicolumn{4}{l}{E$_\gamma$ [keV]}\\
\cline{2-5}
mult. &  30  & 100 & 200 & 500 \\
\hline
E1  &  0.10 & 0.09 & 0.09 & 0.07\\
E2  &  0.07 & 0.07 & 0.11 & 0.08\\
E3  &  0.07 & 0.07 & 0.07 & 0.11\\
E4  &  0.17 & 0.08 & 0.06 & 0.07\\
M1  &  0.10 & 0.13 & 0.10 & 0.08\\
M2  &  0.08 & 0.07 & 0.10 & 0.07\\
M3  &  0.20 & 0.09 & 0.11 & 0.06\\
M4  &  0.22 & 0.10 & 0.07 & 0.07\\
\hline
\end{tabular}
\end{table}
To get more information on the `chemical' uncertainties, we extended
our study to more elements. In particular \dn{87}Fr -- together with
\dn{55}Cs -- belongs to alcaline metals, i.e. those having one 
electron above the noble-gas core. Then \dn{35}Br, \dn{53}I, and
\dn{85}At -- here, one electron lacks to noble-gas-like closed
shells. And finaly \dn{26}Fe, \dn{62}Sm, and \dn{76}Os --
the transition metals. This set covers various configurations
of the valence shell. Moreover, we added the transition energies
of 50 and 300 keV to suppress possibility of some unobserved
changes of the uncertainty.

The results confirmed the conclusions made from the case of \dn{55}Cs.
The changes are of tenths of percent for the iner shells and
increase to tens of percent for the outermost shell. The changes
(uncertainties) of the total ICC are presented in Tab.\ref{t:total}
together with the ranges of ionization used for particular
elements.
\begin{table}[h]
\caption{\protect
Maximum changes of total ICC over all the ionizations given}
\label{t:total}
\begin{tabular}{rrrc}
\hline
\  & \multicolumn{2}{c}{ionization} & change\\
\cline{2-3}
\  & from & to & [\%]\\
\hline
\dn{26}Fe  &  0  & +4  & 0.7\\
\dn{35}Br  &  -1  &  +4  & 0.5\\
\dn{53}I  &  -1  &  +4  & 0.2\\
\dn{55}Cs  &  -1  &  +9  &  0.5\\
\dn{62}Sm  &  0  &  +5 &  0.1\\
\dn{76}Os  &  -1  &  +5 & 0.1\\
\dn{85}At &  -1  &  +4  &  0.1\\
\dn{87}Fr  &  0   &  +8  &  0.2\\
\hline
\end{tabular}
\end{table}
We see that the uncertainty of total ICC due to the chemical
effects decreases with increasing atomic number, \Z. This is
not surprising -- the higher \Z, the more atomic subshells. And
only the lower ones (which are less affected) contribute 
significantly into the total.

{\it \underline{Recommendation}}: If no information on the 
structure of the source exists, the quantities presented in 
Tab.\ref{t:total}
may be used as very conservative estimates of the error.
On the other hand if some information is available, the
present values may be modified (usually decreased) or
neglected.
%%%%%%%%%%%%%%%%%%%%%%%%%%%%%%%%%%%%%%%%%%%%%%%%%%%%%%%%%%%%%
\subsection{Intranuclear conversion}
\label{s:intra}
The intranuclear conversion (also called {\em nuclear structure
effect, penetration effect}) is responsible for that part of ICC
which corresponds to the integral (\ref{e:Tig}). Physically
it is caused by non-zero probability that the electron 
penetrates into the nucleus and the conversion process takes part
there. 

In principle, the nuclear currents $\vec{J}_{nucl}(\vec{R})$ in 
Eq.(\ref{e:Tig}) may be determined from a suitable nuclear model and 
the integral then may be evaluated. Such an ICC, however, would
 be model--dependent. Moreover, the reliability
of nuclear models is not sufficient.
Therefore, two approximations were introduced. The first one, the Rose's
`no penetration' model \cite{Ros58,Gre58}, completely neglects the 
contribution (\ref{e:Tig}). The second one, the Sliv's `surface
current' model \cite{Sli51} assumes that the nuclear transition
 currents are
localized at the nuclear surface which enables the evaluation 
of the integral (\ref{e:Tig}).

Note that the both approaches are approximations only (even when 
each of them is supported by some physical basement). 
A convenient  way how to
manage the contribution of the intranuclear conversion was 
developed by Pauli \cite{Pau67}.
He expressed the ICC involving the contribution (\ref{e:Tig}) as
\beqar
\label{e:alML}
\alpha_i(ML) = \alpha^0_i(ML) \times (1 + b_{1i}\lambda +
    b_{2i}\lambda^2)\\
\alpha_i(EL) = \alpha^0_i(EL) \times (1 + a_{1i}\eta +
    a_{2i}\eta^2 + a_{3i}\eta\xi + a_{4i}\xi + a_{5i}\xi^2)
\label{e:alEL}
\eeqar
where $\alpha_i^0$ is the ICC on the subshell $i$ calculated 
in the `no penetration' model, the formulae for $a$'s and $b$'s 
contain only electronic variables.
All nuclear variables from Eq.(\ref{e:Tig}) are constricted into 
one ($\lambda$)
and two ($\xi$, $\eta$) parameters for the ML and EL multipolarities,
respectively, which, in turn, contain {\em no} electronic
variables. These parameters should be determined by 
comparison of the theoretical and experimental ICC using e.g.
the program \cite{Rys80} or, in principle,  they may be calculated 
in frame of a specific nuclear model.

The possible effect, i.e. the magnitude of the polynomials in 
parentheses in Eqs. (\ref{e:alML},\ref{e:alEL}),
was extensively studied in \cite{Anom,Lis78}.
For non-hindered transitions, the effect -- the change of $\alpha_i^0$
-- varied from negligible \asi 10\up{-2}~\% (for low Z and all
 multipolarities
except M1 and M2) to 10~\% (for high Z and multipolarity M1).
For the hindered transitions, the situation is different. In \cite{Anom},
a lot of various transitions with the hindrance factors $F_W$ 
(related to the Weisskopf single particle estimate) from 10 up to
10\up{16} were studied. It turned out, that the effect fluctuates
between a few percent change up to a change by a {\em factor} of 20.
Unfortunately, there is no universal connection between the hindrance
factor and the magnitude of the effect. As an example, 
the 58 keV E1 transition in $^{180}_{\ 72}$Hf has $F_W$ = 10\up{16}
and the effect reaches the factor 2.5 \cite{Sch67}.
On the other hand, the 84 keV E1 transition 
in $_{\ 91}^{231}$Pa has $F_W$ = 9$\times$10\up{5} only
and the ICC changes up to 20 times \cite{Asa60}.
 (Note that
the change due to the intranuclear conversion is different for
different atomic subshells. When present, the effect changes
the ICC not only for the inner shells but also for the outer
ones \cite{Dra69,Dra71}.)

It is necessary to emphasize here that the intranuclear
conversion itself brings {\em no uncertainty} into the 
{\em theoretical} ICC. The theoretical ICC are calculated using 
one of the above models \cite{Ros58,Sli51}.
In many cases -- when the effect of
nuclear structure is small -- the theoretical ICC enabled
the determination of nuclear characteristics (spins, parities)
fairly well. However, when treating {\em precise} experimental
data or hindered transitions, the intranuclear conversion must be 
taken into account. This means that {\em the ICC, $\alpha_i$, in the 
form of Eqs.(\ref{e:alML},\ref{e:alEL}) must be used instead of
$\alpha_i^0$} and the parameters $\lambda$, $\eta$, $\xi$ 
should be fitted \cite{Rys80} into the experimental data.
The parameters are determined together with their errors 
(standard deviations) but these errors are result of the fit
and they {\em do not} affect the theoretical ICC $\alpha_i^0$.

{\em \underline{Recommendation}}: If there is a discrepancy between
the experimental and theoretical ICC and/or ICC ratios which
cannot be explained by higher multipolarity admixture, always
expect the intranuclear conversion and do fit the parameters
$\lambda$ or $\xi$ and $\eta$.
%%%%%%%%%%%%%%%%%%%%%%%%%%%%%%%%%%%%%%%%%%%%%%%%%%%%%%%%%%%%%%%%
%
\section{Danger of interpolation}
If there is no possibility to calculate ICC directly for the
transition in question, one is forced to use ICC tables.
The tables are also usually used for preliminary 
estimate of effects which can be expected in the detailed
analysis.
The tabulated ICC are always presented in some more or less dense
energy mesh. 
When studying a particular transition,  an interpolation is necessary.

It is generally believed that the dependence of ICC on the transition
energy is close to linear one in the log--log scale (i.e. $\log{ICC}
\sim \log{E_{tr}}$).
If so, the interpolation would make no problems. Unfortunately,
in several cases the situation is not so simple (see e.g. graphs in
\cite{Fir98}). For low-energy
transitions a very non-monotonous  dependence of ICC on energy
was found in \cite{Dra93}.

For the total ICC, the interpolation is strictly forbidden in
the vicinity of the threshold of some atomic shell. As demonstrated
in Fig.\ref{f:total} there is a discontinuity in each point
$E_{tr} = E_{b,i}$ where $E_{b,i}$ is the binding energy of some
atomic subshell. The magnitude of this discontinuity depends
on the transition multipolarity and on the particular subshell.
(E.g. for M1, the discontinuity is greater for $s_{1/2}$ (i.e.
K, L\dn{1}, M\dn{1} etc.) subshells and less for $p_{1/2}$
and $p_{3/2}$ subshells. For E2, it is just opposite.)
Applying an interpolation in such a case would fetch in a
very serious error. It is a pity that  this error
was not avoided in \cite{Fir98} (in graphs of total ICC, appendix F).

\hspace*{-\parindent}
\parbox{11cm}{\epsfxsize=9cm \epsfysize=7cm \epsfbox[20 90 330 410]
{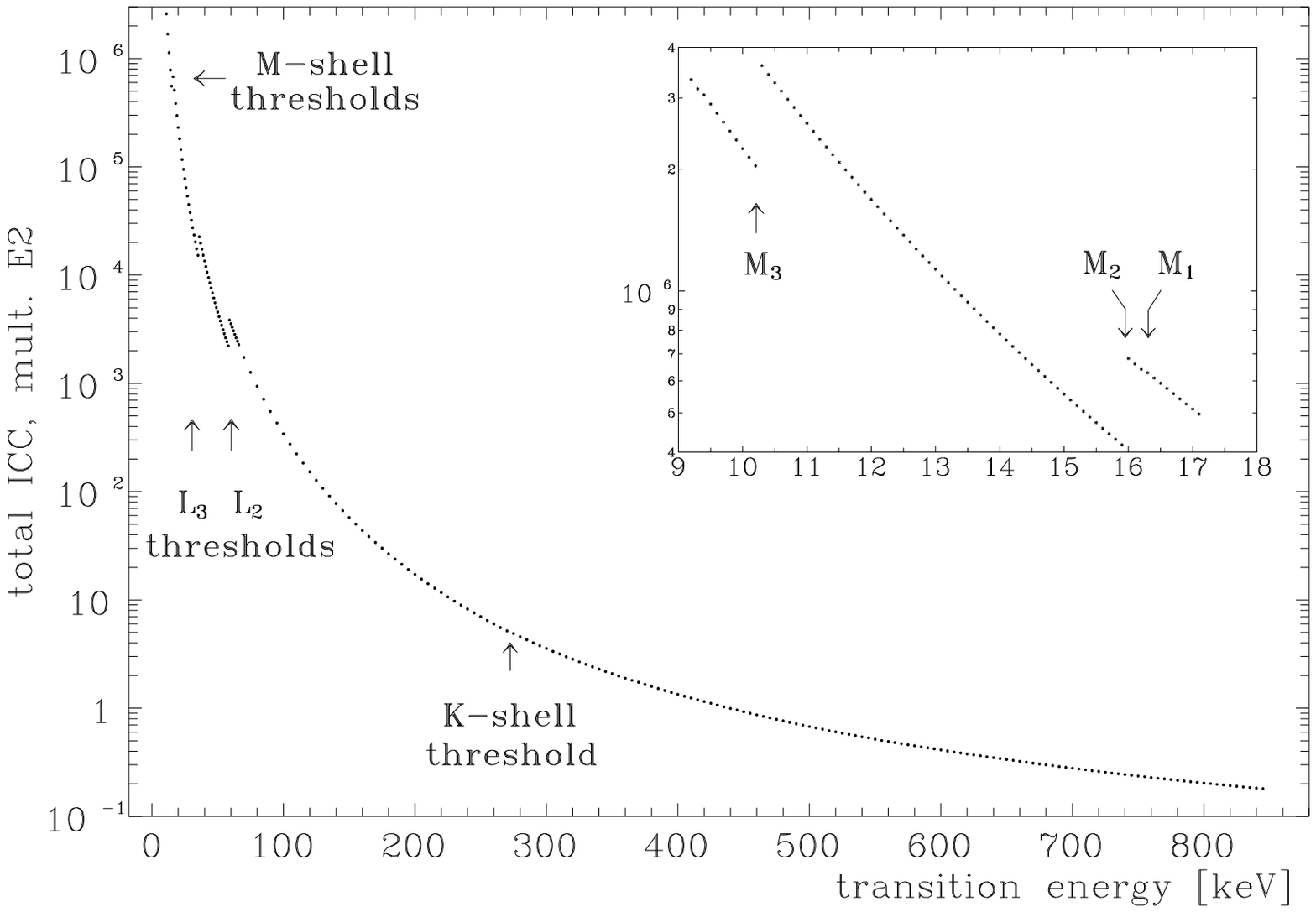}} \vspace{4ex}
\begin{figure}[t]
\protect\caption[Fig3]{Depencence of total ICC on transition energy for 
the E2 transition in element Z=126. (Note that no discontinuity
is observed at the s\dn{1/2}-shell thresholds since the contribution
of these shells into the total ICC is very small in this case.)
}
\label{f:total}
\end{figure}
\section{Conclusions}
We have estimated the uncertainty of the theoretical ICC
relevant to their most frequent  application, i.e. the
spin--parity determination of excited nuclear states.
Until the uncertainties of the experimental ICC and/or
their ratios are better than 10~\%, any of the three 
tables \cite{Roe78,Ban78,Hag68} can be utilized for the comparison
with the experiment and their uncertainties may be neglected.

When the accuracy of the experimental data is of a few percent,
the uncertainty of the theoretical ICC as examined
in this work should be included into the analysis which,
in turn, leads to more realistic error of the final
results (e.g. the multipole mixing parameters). As for the
adding the uncertainties, we recommend to apply
the quadratic addition as in the case of the statistical 
errors. We admit that the mentioned uncertainties are not
of the pure statistical origin but the same argument
as for the statistical errors can be used -- it is improbable
that all deviations assemble into the same direction. 

To circumvent the question of errors caused by the 
atomic model which is not clear enough, the ICC for
the cases of precise experimental data shoud be calculated
directly for the particular transition energy using the
HF atomic model (see Sect.\ref{s:atomic}).

Keeping in mind the possibility of nuclear structure effect
(Sect.\ref{s:intra})
in hindered transitions, more than one experimental
quantity (ICC and/or ICC ratio) shoud be fetched into
analysis whenever possible.

The interpolation in the tables of total ICC should be avoided.
This is due to discontinuities of total ICC at the points of
particular subshell's threshold. For lower transition energies,
even the dependence of subshell ICC on the energy becomes
non-monotonous and the interpolation must be done
with a great care.

\vspace{10mm}\ 

{\it This work was supported by the Grant Agency of the Czech Republic
under the contract No. 202/00/1625.}
\end{document}